      [1999/12/01 v1.4c Il Nuovo Cimento]
\documentclass{cimento}
\usepackage{epsfig}
\usepackage{graphicx,psfrag}
\usepackage{txfonts}
\title{INTEGRAL and XMM-Newton observations of the low-luminosity and X-ray rich burst GRB 040223}
\author{S.~McGlynn\from{ins:1}, S.~McBreen\from{ins:2},
L.~Hanlon\from{ins:1}, B.~McBreen\from{ins:1}, S.
Foley\from{ins:1}, L. Moran\from{ins:3}, R.~Preece\from{ins:4},
A.~von~Kienlin\from{ins:5} \atque O.R. Williams\from{ins:6}}
\instlist{\inst{ins:1} Department of Experimental Physics,
University College, Dublin 4, Ireland
\inst{ins:2} Astrophysics
Missions Division, RSSD of ESA, ESTEC, Noordwijk, The Netherlands
  \inst{ins:3} 
  School of Physics and Astronomy, University of Southampton, Southampton, 
SO15 3EZ, UK
  \inst{ins:4} The University of Alabama in Huntsville, AL 35899,
  USA
\inst{ins:5}Max-Planck-Institut f{\"u}r extraterrestrische Physik,
Giessenbachstrasse 85748, Garching, Germany \inst{ins:6} Science
Operations and Data Systems Division of ESA/ESTEC, SCI-SDG, 2200
AG Noordwijk, The Netherlands}
 \PACSes{\PACSit{98.70.Rz}{$\gamma$-ray sources: $\gamma$-ray bursts}
\PACSit{95.55.Ka}{X-ray and $\gamma$-ray telescopes and
instrumentation.}}

\begin{document}

\maketitle

\begin{abstract}
GRB 040223 was observed by INTEGRAL and XMM-Newton.  GRB 040223
has a peak flux of (1.6 $\pm 0.13) \times 10^{-8}$ ergs cm$^{-2}$
s$^{-1}$, a fluence of (4.4 $\pm$ 0.4) $\times$ 10$^{-7}$ ergs
cm$^{-2}$ and a steep photon power law index of -2.3 $\pm$ 0.2, in
the energy range 20--200~keV.  The steep spectrum implies it is
an x-ray rich GRB with 
emission up to 200 keV and \(E_{\rm peak} <
20\) keV. If \(E_{\rm peak}\) is $<$ 10~keV, it would qualify as an
x-ray flash with high energy emission.  The x-ray data has a
spectral index \(\beta_{x} = -1.7 \pm 0.2\), a temporal decay of
t$^{-0.75 \pm 0.25}$ and a large column density of \(1.8 \times
10^{22}\) cm\(^{-2}\). The luminosity-lag relationship was used to
obtain a redshift  (z = 0.1$^{+0.04}_{-0.02}$ ). The isotropic energy
radiated in $\gamma$-rays and x-ray luminosity after 10 hours are
factors of about 1000 and 100 less than classical GRBs.  GRB 040223 is
consistent with the extrapolation of the Amati relation into
the region that includes XRF 030723 and XRF 020903.
\end{abstract}

\section{Introduction}
The prompt emission from GRBs and the afterglow give valuable
information on the radiation processes and the environment.  There
is much evidence that the $\gamma$-radiation originates from
dissipative processes in a relativistically expanding plasma wind
with either shocks or in magnetic reconnection.  In addition to
GRBs, x-ray flashes (XRF) have been identified as a class of soft
bursts that are very similar to GRBs [e.g. 2].  There seems to be
a continuum of spectral properties for XRFs, x-ray rich GRBs and
classical GRBs and it is probable that they have a similar
origin~\cite{ref:sakamoto2004}.

ESA's International Gamma-Ray Astrophysics Laboratory \textit{INTEGRAL} 
(\cite{ref:winkler2003}), launched in October 2002, 
is composed of two main coded-mask telescopes, an imager IBIS 
\cite{ref:ubertini2003}
 and a spectrometer SPI 
 \cite{ved2003})
coupled with two monitors, one in the X-ray band and in the optical band. 
The two main instruments on board have coded masks, a
wide field of view and cover a wide energy range (15~ keV- 8~MeV).
The IBIS instrument consists of two independent solid state arrays ISGRI 
and PICsIT, optimised for low and high energies repsectively.
The ISGRI detector of IBIS consists of an array of 128 x 128 
CdTe crystals sensitive to lower-energy gamma-rays (\cite{leb:2003})
and is most sensitive between 15--300~keV. The mask is located
3.4~m above the detector plane. 
The SPI detector plane consists of an array of 19 germanium detectors
surrounded by an active anticoincidence shield of BGO  and imaging
capabilities are achieved with a tungsten shield located 1.7~m 
from the Ge array. SPI is sensitive from $\sim$20~keV --8~MeV.

\textit{INTEGRAL} has a burst alert system called IBAS (\textit{INTEGRAL} Burst Alert System, \cite{ref:mereghetti2003}).
\textit{INTEGRAL}  has detected and localised 21 GRBs to an accuracy of a few arcminutes using
IBAS. Most of these bursts are very
weak~e.g.~\cite{ref:moran2004}.  SPI light curves and spectra of some of
the bursts are publicly available~\cite{ref:http}.  Some bursts
such as GRB 030329 are less luminous in $\gamma$-rays than the
standard value of \(\sim 10^{51}\) ergs~\cite{ref:bloom2003}. The
bursts GRB 980425 and GRB 031203 are in a separate class of
sub-energetic
events~\cite{ref:sazonov2004,ref:soderberg2004,ref:watson2004}.

Here we report on observations of the prompt emission of GRB 040223 with 
\textit{INTEGRAL} and the afterglow with XMM-Newton.
The afterglow was not detected in the
optical or near infrared~\cite{ref:tagliaferri} probably because
of the large absorption associated with the high column density ($\S$\ref{sect:analysis}).
No radio emission was detected from the afterglow about one day
after the burst~\cite{ref:soderberg}.

\section{Data analysis and results}
\label{sect:analysis}
GRB 040223 was detected by the INTEGRAL burst alert system
IBAS~\cite{ref:mereghetti2003} and the location rapidly
distributed~ \cite{ref:gott}.  The IBIS light curve is given in
Fig.~1a and does not include two very weak emission pulses at
about -110 s and -180 s. The weak emission was also detected with
SPI.  GRB 040223 is in the long duration class with a well
resolved pulse (Fig 1~a). The IBIS light curve was denoised with a wavelet
analysis~\cite{ref:quilligan2002} and the risetime, fall time and
FWHM of the pulse are 19~s, 22~s, and 13~s respectively. 
GRB 040223 fits well with the expected trends from previous 
analyses \cite{ref:mcbreen2002} of BATSE bursts consisting
of only a few pulses (Fig~2).
 The IBIS data was divided into two energy
channels i.e. 25--50~keV and 100--300~keV. The cross-correlation
analysis~\cite{ref:norris2002,ref:schaefer2004} was performed
between the two channels and the lag was determined to be 2.2$\pm 0.3$ ~s which is longer than observed in most
GRBs~\cite{ref:norris2002}.
\begin{figure}[t]
\psfrag{Counts}{\tiny Counts}
\psfrag{Time (s)}{\tiny \\ \\ Time (s)}
\psfrag{(a)}{\footnotesize (a)}
\psfrag{(b)}{\footnotesize (b)}
\psfrag{(c)}{\footnotesize(c)}

\begin{minipage}[c]{.5\textwidth}
\psfrag{Residuals}{\footnotesize Residuals}

\vspace{-0.25cm}
\includegraphics[width=\textwidth]{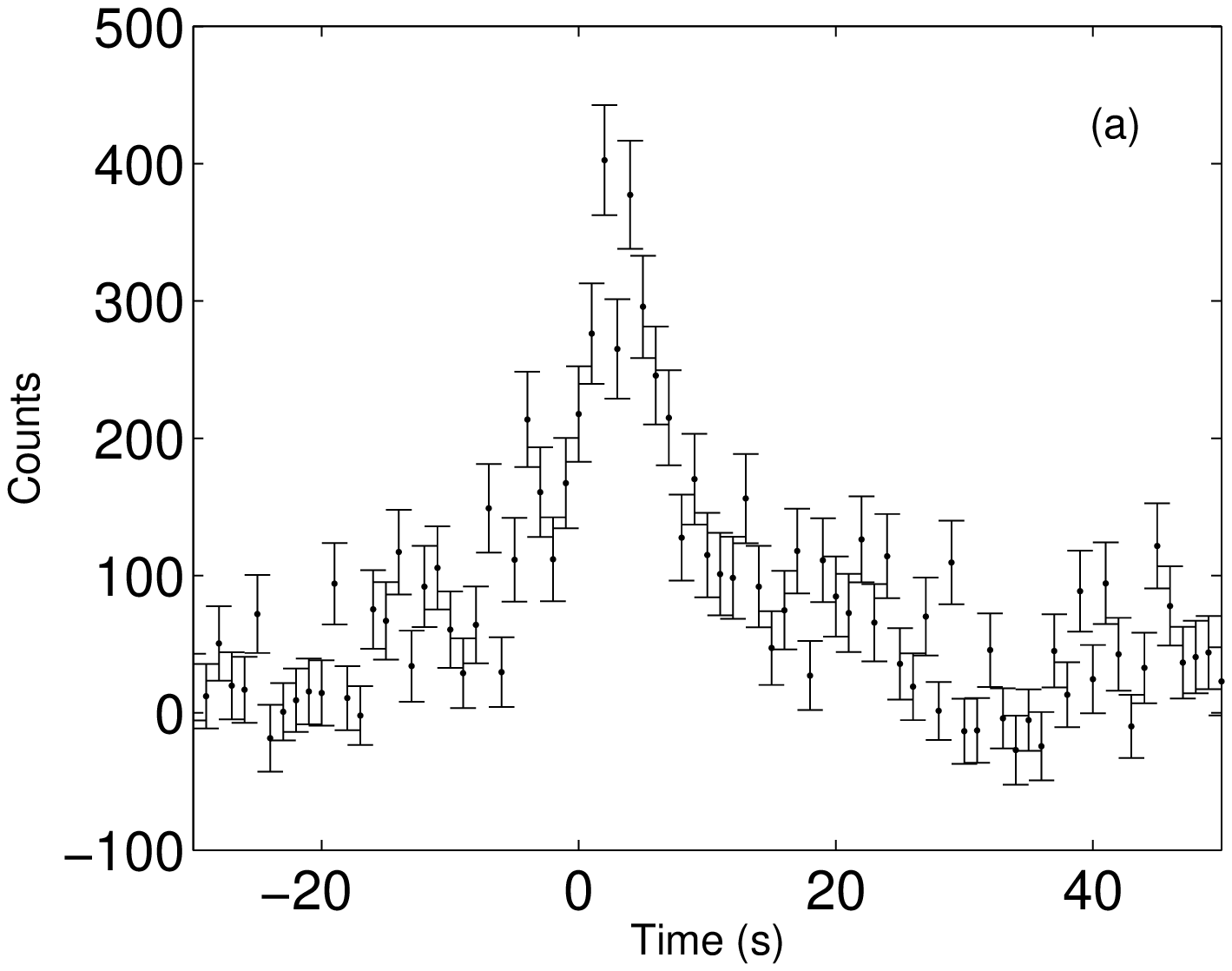}
\end{minipage}%
\begin{minipage}[c]{.5\textwidth}
\includegraphics[width=\textwidth]{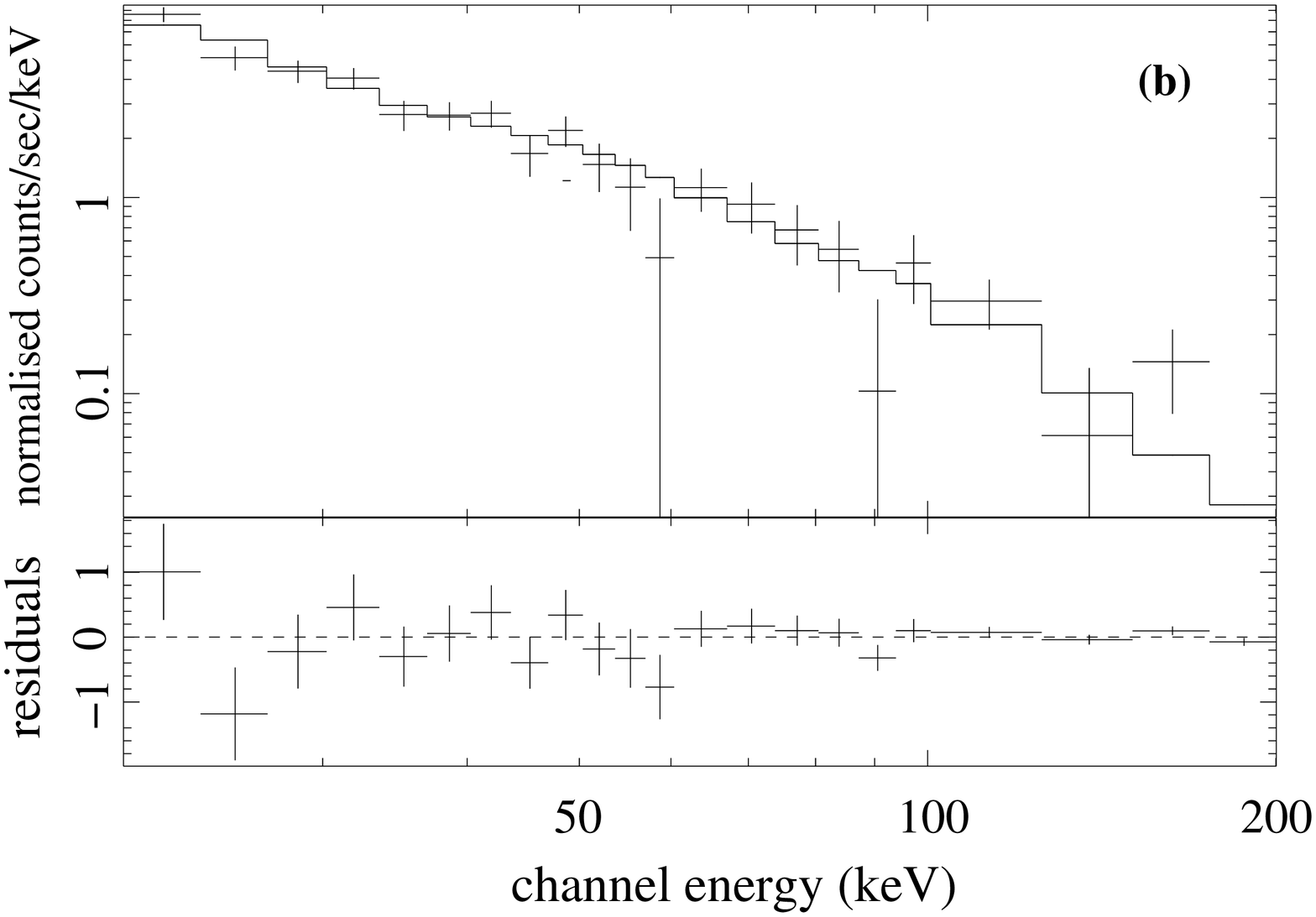}

\end{minipage}

 \caption{{\bf a)}~IBIS lightcurve of GRB 040223 in the energy range 15--200~keV
 and zero time is the IBAS trigger at 13:28:10 UTC.  {\bf b)}~IBIS spectrum of GRB 040223 fit by a power law model from 20--200~keV.
Upper panel: data and best fit model. Lower panel: residuals
between the data and the folded model.
}
\end{figure}

\begin{figure}[t]
\psfrag{Counts}{\tiny Counts}
\psfrag{Time (s)}{\tiny \\ \\ Time (s)}
\psfrag{(a)}{\footnotesize (a)}
\psfrag{(b)}{\footnotesize (b)}
\psfrag{(c)}{\footnotesize(c)}

\includegraphics[width=\textwidth]{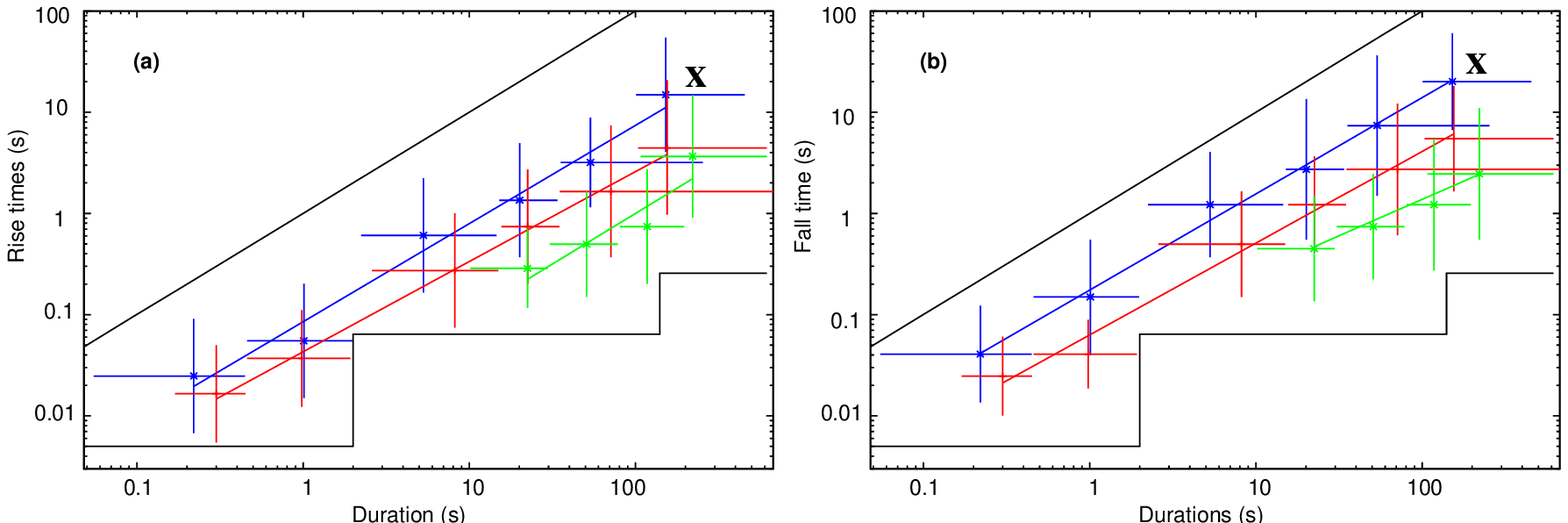}


 \caption{
 Timing
 diagrams in BATSE GRBs and GRB 040223. The median values for  {\bf a)} rise time 
,  {\bf b)} fall time  
are plotted versus
duration T$_{90}$ for GRBs in three categories i.e. 1 $\leq$ N
$\leq$ 3 (blue), 4 $\leq$ N $\leq$ 12 (red) and N $>$ 12 (green)
where N is the number of pulses detected in a burst.
The crosses signify the range covered in T90 and e$^{\mu \pm
\sigma}$ for the lognormal distribution which includes 16\% to
84\% of the pulse values for that T$_{90}$ bin.  The upper
diagonal line is the limit where the pulse parameter is equal to
T$_{90}$ and the lower lines by the limited time resolution i.e.
5 ms, 64 ms and 256 ms.  The \textbf{X} marks the values for GRB~040223.
}
\end{figure}

\begin{figure}[t]
\psfrag{Counts}{\tiny Counts}
\psfrag{Time (s)}{\tiny \\ \\ Time (s)}
\psfrag{(c)}{\footnotesize(a)}
\psfrag{(b)}{\footnotesize(b)}

\begin{minipage}[c]{0.45\textwidth}
\includegraphics[width=\textwidth]{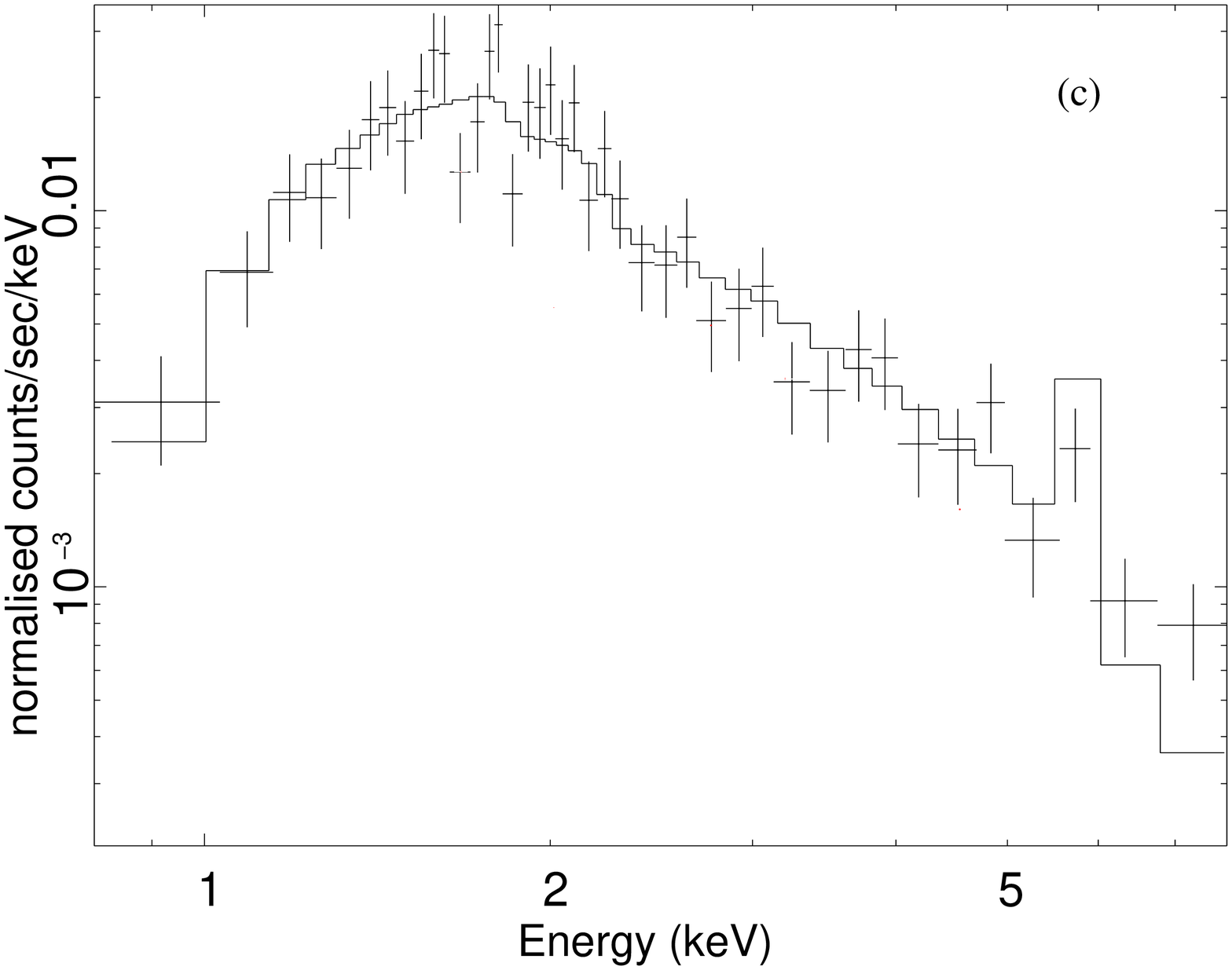}
\end{minipage}
\begin{minipage}[c]{0.53\textwidth}
\includegraphics[width=\textwidth]{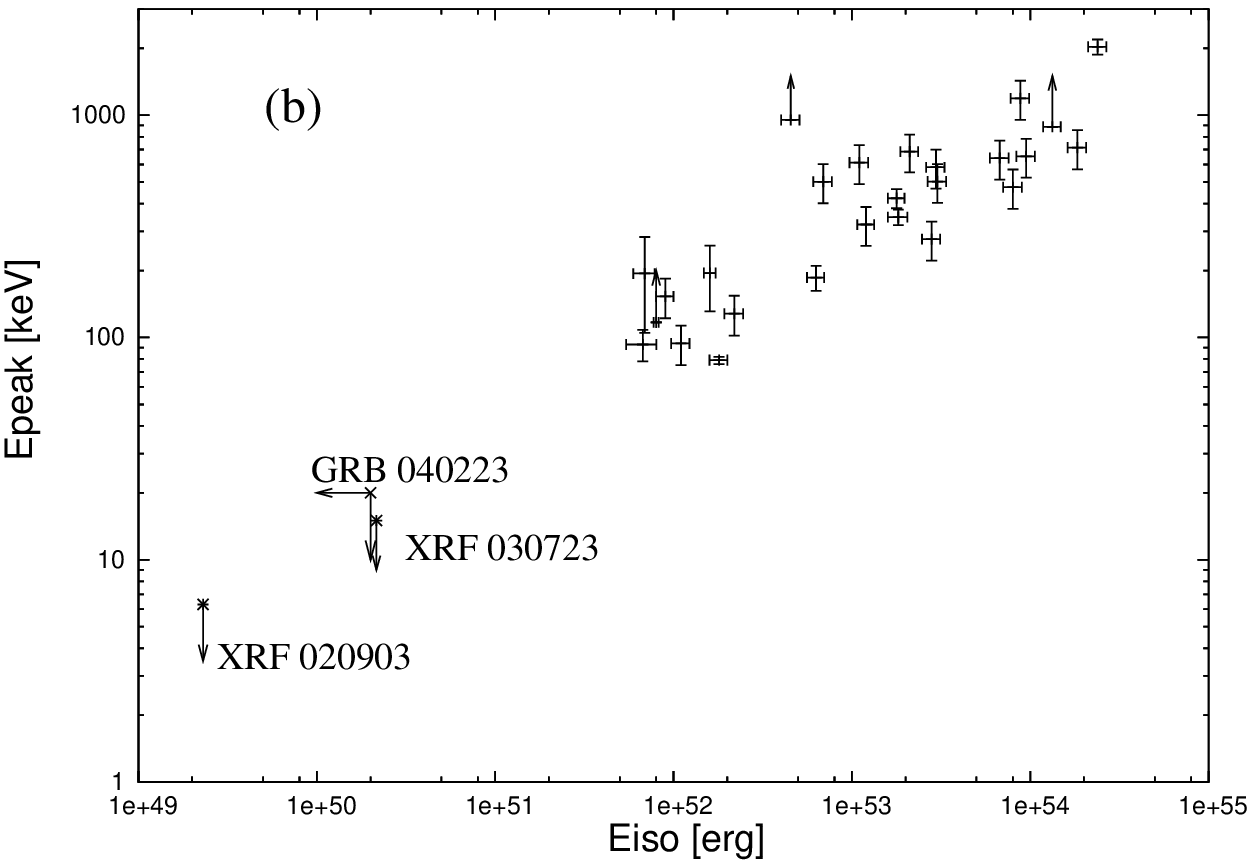}
\end{minipage}
 \caption{{\bf a)}\,EPIC spectrum of the GRB 040223 afterglow
and its best fit absorbed power law model. The 
 data points shown
refer to PN and agree with the  MOS data. 
{\bf b)} Data from a sample of GRBs used 
to obtain a relationship between E$_{\rm peak}$ and E$_{\rm ISO}$
\cite{ref:amati2002,ref:ghirlanda2004} .
GRB~040223 lies in the region occupied by the two x-ray flashes.}
\end{figure}

The IBIS spectral analysis was performed using the method
described for GRB 040106~\cite{ref:moran2004,ref:filliatre2005}.
The IBIS data, in the range 20 to 200~keV, is well fit by a single
power law with photon index -2.3 $\pm$ 0.2 with a reduced
$\chi^{2}$ of 1.01 and 20 degrees of freedom (dof) (Fig.~1b).  The
quoted errors are at the 90\% confidence level.  The peak flux is
(1.6 $\pm 0.13) \times$ 10$^{-8}$ ergs cm$^{-2}$ s$^{-1}$ in the
brightest second and fluence (4.4 $\pm 0.4) \times$ 10$^{-7}$ ergs
cm$^{-2}$. The spectral results obtained with SPI are consistent
with IBIS and $\gamma$-ray emission was detected  to 200~keV.

XMM-Newton observed the location of the GRB for 42 ks starting
only 18 ks after the burst.  A fading x-ray source was detected at
RA = 16$^{h}$ 39$^{m}$ 30.17$^{s}$, Dec =
-41$^{0}$ 55' 59.7'' within the IBAS error region. The temporal
decay of the x-ray afterglow (F$_{\nu}(t) \propto t^{-\delta}$)
was fit by a power law with index $\delta$ = -0.7 $\pm$
0.25 by Gendre et al. (2004)~\cite{ref:gendre2004}.  Only 17 ks of the light curve was
used in the analysis because of contamination. Our analysis is consistent with this result.

We obtained afterglow spectra from the PN and MOS
Cameras~\cite{ref:turner2001} 
 after standard data
screening.  The three spectra in the energy range 0.2 --10~keV
were well fit by a power law 
\(F_{\nu} \propto \nu^{-\beta_{x}}\)  where the spectral index
\(\beta_{x} = 1.7 \pm 0.2\) with reduced $\chi^{2}$ of 1.29 for
111 dof (Fig.~3a). The absorption
column density has a high value of \(N_{H} = 1.8 \times 10^{22}\)
cm$^{-2}$ and exceeds the galactic value in this direction of \(6
\times 10^{21}\) cm$^{-2}$.  
\section{Discussion}
It is suspected that the nearest and most frequent bursts have
been missed because they are either intrinsically sub-energetic or
off-axis or both~\cite{ref:norris2002,ref:quilligan2002}.  In an
extensive luminosity-lag study~\cite{ref:norris2002} a sub-sample
of bursts was identified that had few wide pulses, long lags, soft
spectra and a Log N--Log S distribution that approximates to a
-$\frac{3}{2}$ power law. GRB 040223 consists of a slow pulse with
a long lag of 2.2$\pm 0.3$ seconds,  lies near the
supergalactic plane with coordinates 179$^{0}$, 24$^{0}$ and seems
to be a member of the sub-sample.

There are no direct measurements of the redshift to GRB 040223 and
indirect, model dependent, distance indicators must be used.  The
luminosity-lag relationship was modified~\cite{ref:norris2002} to
include GRB 980425 and the peak luminosity is given by
\[L = 7.8 \times 10^{53} \left(\frac{\tau_{lag}}{0.1\,
s}\right)^{-4.7}\] \noindent for \(\tau_{lag} > 0.35\) s.  The
peak luminosity is \(3.8^{+3.8}_{-1.7} \times 10^{47}\) ergs
s$^{-1}$ for the lag of 2.2$\pm0.3$~s. The source redshift
is z = 0.1$^{+0.04}_{-0.02}$ when the peak flux of 1.6 $\pm
0.13 \times 10^{-8}$ ergs cm$^{-2}$ s$^{-1}$ is combined with the
peak luminosity.  The  luminosity distance is 460$^{+200}_{-130}$~Mpc
using standard cosmology.
The fluence gives a total isotropic $\gamma$-ray
luminosity (E$_{\rm ISO}$) of approximately \(10^{49}\) ergs which is
about three orders of magnitude less than classical GRBs. GRB
040223 is sub-luminous in $\gamma$-rays by a large factor.

The $\gamma$-ray luminosity may also be obtained using the   Amati relation~\cite{ref:amati2002}. The measured photon index
is well outside the normal range~\cite{ref:preece2000} for
$\alpha$ of $\frac{-3}{2}$ to $\frac{-2}{3}$ and the value of -2.3
$\pm$ 0.2 is remarkably similar to the spectral index $\beta$
above the break energy (E$_{o}$) that has a value typically
between -2 and -2.5. The slope is steeper than the value of
-1.9 reported for the weak INTEGRAL burst GRB
040403~\cite{ref:mereghetti2004}. GRB 010213 also has a very soft
spectrum with a photon index of -2.14 and a low value for E$_{o}$
of 4 keV~\cite{ref:barraud2003}. If it is assumed that $\beta$ =
-2.3 $\pm$ 0.2 then E$_{o}$ must be less than 20~keV to be outside
the IBIS energy range.  The peak energy E$_{\rm peak}$, is given by (2
+ $\alpha$) E$_{o}$ and hence E$_{\rm peak}$ = E$_{o}/2$, for a
value of $\alpha$ = $-$1.5. 
Using the Amati
relationship~\cite{ref:ghirlanda2004}, the value of E$_{\rm ISO}$ is
\(< 2 \times 10^{50}\) ergs assuming that the rest frame value of
E$_{\rm peak}$ is $<$ 20~keV.  The two indirect distance indicators
yield consistent results and show that GRB 040223 lies on or near
the extrapolation of the Amati relation from classical GRBs to
include XRF 030723 and XRF 020903 (Fig.~3b). 
The Yonetoku relation \cite{ref:Yonetoku} between peak luminosity and E$_{\rm peak}$ also yields results consistent with the above conclusions.
 GRB 040223 is an x-ray rich burst with a
low value of E$_{\rm peak}$ and it would qualify as an XRF if E$_{\rm peak}$
is $<$ 10~keV~\cite{ref:barraud2003,ref:granot}.  

The x-ray flux after 10 hours is 2.4 $\pm$ 0.4 $\times$ 10$^{-13}$
ergs cm$^{-2}$ s$^{-1}$ in the 2-10~keV region.  The x-ray
luminosity of GRB 040223 is \(6 \times 10^{42}\) ergs s$^{-1}$ and
is about 2 orders of magnitude fainter than observed from
classical GRBs~\cite{ref:bloom2003}.  It is interesting to note
that the x-ray and $\gamma$-ray luminosities of GRB 040223 are
only a factor of 2 and 5 weaker than the sub-energetic burst GRB
031203~\cite{ref:sazonov2004,ref:soderberg2004}. The major
difference between the two bursts is that GRB 031203 has $\alpha$
= 1.65 $\pm$ 0.1 and E$_{o}
>$ 180~keV~\cite{ref:sazonov2004} whereas GRB 040223 has a steeper index of -2.3
$\pm$ 0.2 and E$_{o} <$ 20~keV.

The x-ray afterglow of GRB 040223 has a slow decline \((\delta =
-0.75 \pm 0.25)\) that is similar to slow declines of many
GRBs~\cite{ref:feroci2001,ref:fynbo2004} including \(\delta = -1.0
\pm 0.1\) for XRF 030723~\cite{ref:butler} and \(\delta = -1.1\)
for XRF 020903~\cite{ref:sakamoto2004,ref:soderberg1}. The x-ray
and $\gamma$-ray luminosities of GRB 040223 and XRF 030723 are
comparable.  It seems likely that GRBs are viewed from inside
the jet whereas XRFs are viewed from the side and have lower
luminosities, longer lags and slower and fewer
pulses~\cite{ref:granot,ref:mcbreen2002,ref:norris2002,ref:quilligan2002}.
GRB~040223 seems to lie at low redshift and in a region where
a modification of the luminosity function appears to be necessary 
to account for the number of observed bursts \cite{ref:coward2005,ref:guetta2004}.

\end{document}